\documentclass[]{spie}

\usepackage{amsmath,amsfonts,amssymb}
\usepackage{graphicx}
\usepackage{csquotes}
\usepackage{multirow}
\usepackage{array}
\usepackage{caption}
\usepackage[colorlinks=true, allcolors=blue]{hyperref}
\title{The second Little UV Camera (LUVCam2) mission}

\author[a]{Emily Hu}
\author[a, b]{Suresh Sivanandam}
\author[b]{Albert Lau}
\author[b]{Coby Silayan}
\author[b]{Mark Barnet}
\author[b]{Shaojie Chen}
\author[b]{Gavin Hay}
\author[b]{Nadim Hatoum}
\author[c]{Aaron Tohuvavohu}
\author[d]{Norbert Werner}
\author[d]{Marianna Dafčíková}
\author[d]{Filip Münz}
\author[b]{Patrick Nkwari}
\author[d]{Jakub Řípa}
\author[e]{Andras Pal}

\affil[a]{David A. Dunlap Department of Astronomy \& Astrophysics, University of Toronto, 50 St. George Street, Toronto, Canada}
\affil[b]{Dunlap Institute for Astronomy and Astrophysics, University of Toronto, 50 St. George Street, Toronto, Canada}
\affil[c]{Cosmic Frontier, USA}
\affil[d]{Department of Theoretical Physics and Astrophysics, Faculty of Science, Masaryk University, Kotlářská 267/2, 611 37 Brno, Czech Republic}
\affil[e]{Konkoly Observatory, Research Centre for Astronomy and Earth Sciences, Budapest, Hungary}

\authorinfo{Further author information: (Send correspondence to E.H.)\\E.H.: E-mail: ejw.hu@mail.utoronto.ca}

\begin{document} 
\maketitle

\begin{abstract}
The second Little UltraViolet Camera (LUVCam2) is a 80\,mm-diameter dual UV-optical space telescope that will observe the UV flares around low-mass stars, with planned flight as a $\sim$1.4U payload aboard the 3U CubeSat BRNOsat in late 2027. We present the concept of operations, full system architecture, optomechanical, and electrical design for LUVCam2. LUVCam2 is the second technology demonstration for the LUVCam electronics board, with the first LUVCam (LUVCam1) still operating in low-Earth orbit (launched July 2024). The main upgrade of LUVCam2 is the use of the backside-illuminated GSENSE4040BSI CMOS sensor, promising significant improvement to UV quantum efficiency over the frontside-illuminated model used in LUVCam1. Other upgrades to LUVCam2 include the incorporation of the LUVCam on-board computer onto the readout electronics, an integrated 1\,W heater and thermal control system, and novel operational capability for LUVCam2 to perform on-sensor star-tracking using optical images of background stars. 
\end{abstract}


\section{INTRODUCTION}
\label{sec:intro}

\begin{figure}[ht!]
    \centering
    \includegraphics[width=0.33\linewidth]{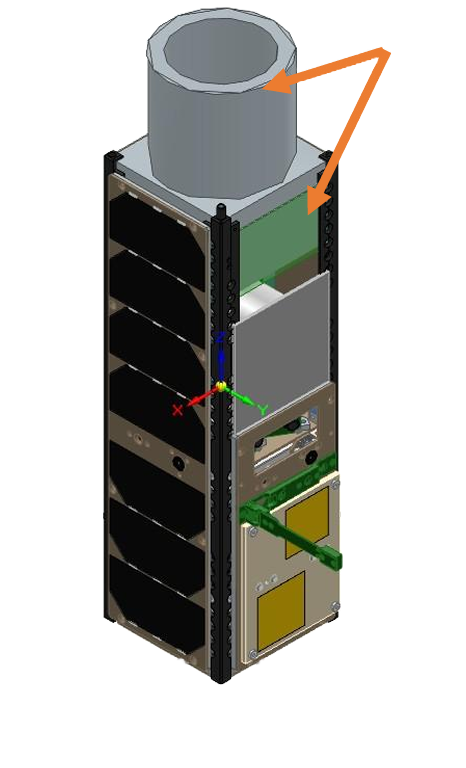}
    \includegraphics[width=0.6\linewidth]{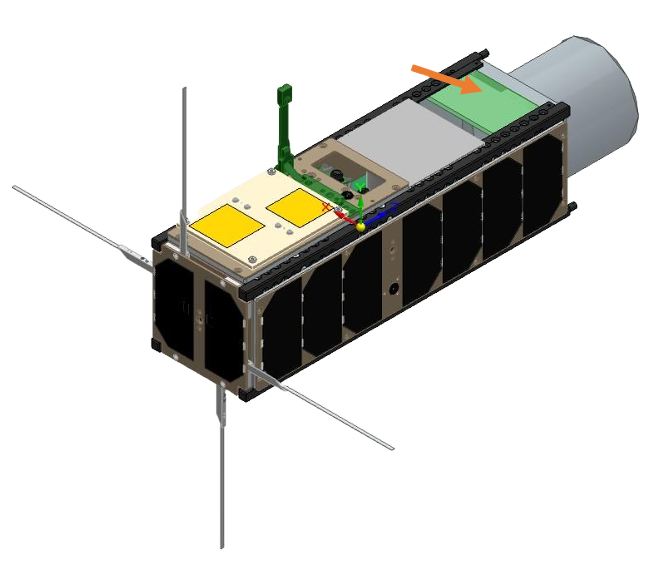}
    \caption{Two views of the payload allocation for LUVCam2 (\enquote{tuna can} and shaded green), labelled with orange arrows, aboard the 3U CubeSat BRNOsat.}
    \label{fig:tuna_can}
\end{figure}

The second Little UltraViolet Camera (LUVCam2) mission is a compact, dual-band UV-optical telescope that will observe the flaring activity of M-dwarf stars. The mission serves as both a technological demonstration of its upgraded general CMOS image sensor platform, and also as a low-cost space mission for assessing exoplanet habitability around M-dwarfs. For this iteration of the Little UltraViolet Camera (LUVCam), the CMOS image sensor platform will integrate the Gpixel GSENSE4040BSI image sensor, the most sensitive UV sensor on the market at a commercial price point at the time of development\cite{Gill_2022}. The payload consists of the LUVCam2 electronics board, the GSENSE4040BSI image sensor, a 1\,W heater and thermal control system, the optical tube assembly, and the mechanical housing. LUVCam2 will be integrated into the \enquote{tuna can} inclusion of BRNOsat (Figure \ref{fig:tuna_can}), a 3U CubeSat Czech mission. With the expected launch of BRNOsat in late 2027, LUVCam2 will be flown in a low-Earth polar Sun-synchronous orbit alongside other on-board technology demonstrations, including an experiment with lubricants and labyrinth seals and a ``smart" thermal switch from Brno University of Technology.  

The optical tube assembly on LUVCam2 is an 80\,mm-diameter catadioptric telescope with a $4^\circ\times4^\circ$ field of view and f/3.7. The aperture size is chosen such that it maximizes the available payload volume aboard BRNOsat. The split-field filter is over the LUV band (LUVCam2 UV; 250--350\,nm) and optical band (SDSS g-band; 410--550\,nm) to enable colour analysis of M-dwarf flares and active guiding on background stars.

The LUVCam board is a low-cost, general purpose CMOS image sensor platform that enables seamless integration of commercial off-the-shelf CMOS sensors into space camera systems. By rapidly iterating its development over several CubeSat missions, LUVCam has quickly raised its Technology Readiness Level in preparation for its integration as the primary science camera on the ESA Quick UltraVIolet Kilonova Surveyor (QUVIK\cite{Werner_2022, Daniel_2024, Werner_2024}; planned launch 2030). The first Little UltraViolet Camera (LUVCam1) mission launched in July 2024 aboard the 2U CubeSat GRBBeta and integrated the GSENSE4040FSI image sensor. While LUVCam1 is actively performing imaging passes in-orbit, technological development for the second Little UltraViolet Camera (LUVCam2) is currently underway. LUVCam2 will build on the operational success of LUVCam1, improving on sensitivity performance, thermal stability, and imaging fidelity of on-board optics. 

We provide an overview of the LUVCam2 payload design and report on the current development status. Section \ref{sec:science} introduces the science case and operational capabilities of the LUVCam2 mission, Section \ref{sec:payload} overviews LUVCam2's system architecture, Section \ref{sec:current} reports on key development milestones achieved thus far, and Section \ref{sec:conclusion} summarizes and concludes.

\section{SCIENCE \& OPERATIONS} \label{sec:science}

\begin{figure}[ht!]
    \centering
    \includegraphics[width=0.7\linewidth]{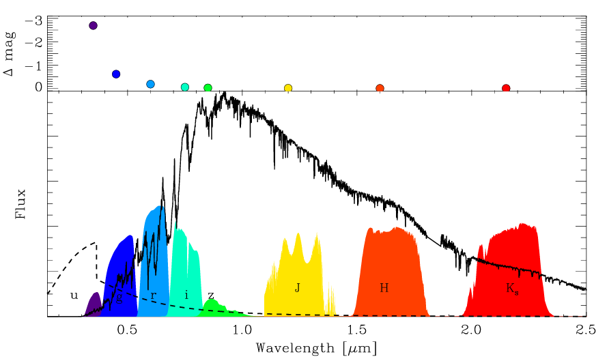}
    \caption{The two-component flare model (dashed line) by Davenport et al. (2012)\cite{Davenport_2012} shown against a quiescent M3-dwarf optical-NIR stellar SED (black line). The ugrizJHKs photometric filter passbands are shown for reference. Figure originally produced in Davenport et al. (2012)\cite{Davenport_2012}.}
    \label{fig:flares}
\end{figure}

M-dwarf stars are the most abundant stellar type \cite{Salpeter_1955, Reid_1997} and known host of exoplanets in our Universe \cite{Howard_2012, Mulders_2015}, drawing intense interest as potential hosts of habitable exoplanets. However, such low-mass stars have fully convective interiors that drive strong magnetic activity at their surfaces, producing frequent sun-spots and flares. In the presence of intense stellar activity, an exoplanets' atmosphere is quickly stripped and cannot develop the chemical signatures for supporting life on the planet's surface \cite{Rimmer_2018}, causing abiogenesis. Complexly, the pre-biotic chemical processes necessary for forming life are also most sensitive in the UV\cite{Ranjan_2017, Rimmer_2021}, where M-dwarf flares are the strongest\cite{Mamonova_2026} (Figure \ref{fig:flares}). Few observations have been made of M-dwarf flares in the UV band due to inaccessibility from ground-based observatories, the oversubscription of modern space telescopes, and the limited dynamic range of traditional space UV sensors. For instance, micro-channel plate (MCP) detectors are most commonly used by UV surveyors for their storied flight heritage. However, these surveyors are required to establish bright-object protection against M-dwarf flares, as intense UV overexposure to MCPs will damage them irreparably\cite{Payne_2026}.

LUVCam2 shall be dedicated towards studying the UV and optical flares on M-dwarfs, as an indirect assessment for habitability on their orbiting exoplanets. 

\subsection{Target Requirements}

LUVCam2 will provide the UV-counterparts to a number of recent flaring M-dwarf observations in the optical band\cite{Gunther_2020}, grounding theoretical optical-UV flare models with direct observations\cite{Jackman_2022}. A select set of bright, nearby M-dwarfs will be determined for LUVCam2's observational program. 

\begin{figure}[ht!]
    \centering
    \includegraphics[width=0.8\linewidth]{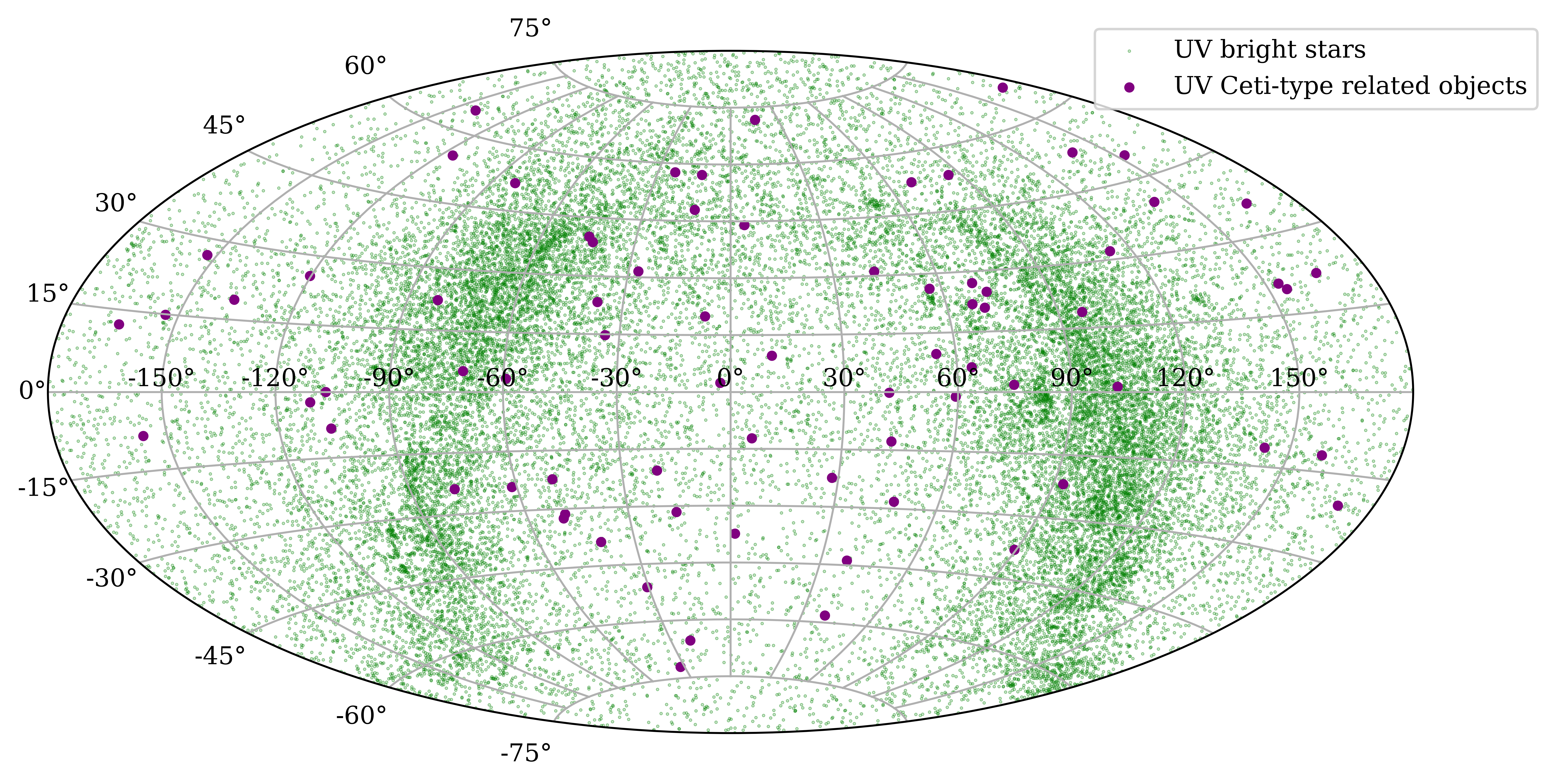}
    \caption{Sky distribution of UV bright stars\cite{Boksenberg_1973} and UV-Ceti type stars\cite{Gershberg_1999} in the Galactic frame.}
    \label{fig:uvceti}
\end{figure}

Since M-dwarfs are faint in the UV-optical bands in their quiescent state\cite{France_2013}, we limit the observations by LUVCam2 to M-dwarfs in the solar vicinity. The Catalogue of UV-Ceti type flare stars (that are mainly late-type K- and M- dwarfs) and related objects in the solar vicinity\cite{Gershberg_1999} is referenced to characterize the sky distribution of known sources in Figure \ref{fig:uvceti}. In addition to imaging bright M-dwarfs, LUVCam2 will also prioritize candidates with known exoplanets. The final selection of M-dwarfs will reflect a combination of targets with high UV stellar activity, confirmed exoplanet hosts, and objects historically well-studied in complimentary optical-NIR bands.

Recurrent photometric observations of targets will be performed such that the target is sufficiently sampled over the duration of its flare; UV Ceti-type stars can flare as often as 1 hour$^{-1}$ and for tens to thousands of seconds\cite{Gershberg_1972, Ishida_1990}. For these frequently flaring stars, LUVCam2 will take several images of the star within a single flare period, to ensure that the flare event is detected alongside multiple baseline exposures of the star in its quiescent state. The imaging frequency will be limited by LUVCam2 readout electronics and availability of on-board memory. To perform differential photometry in the event of a flare observation, we will simultaneously image UV Ceti-type targets alongside UV-bright objects within the background field. This requires a sufficiently dense field of UV-bright sources in LUVCam2's field of view, thus setting a minimum requirement on the field of view and sensitivity for the optical design.

\subsection{Pointing Stability}

Preliminary science results from LUVCam1 (currently in-orbit, launched July 2024) show that the experimental the Attitude Determination and Control System (ADCS) aboard GRBBeta severely limits the camera's ability to obtain stable, high-resolution science images. LUVCam2 will fly aboard BRNOsat, which will integrate a similar, experimental ADCS. 

To enable stable pointing for scientific imaging, LUVCam2 will perform active, on-sensor star-tracking with its images. The on-board FPGA will detect and compute star centroids to establish a guiding feedback loop with the BRNOsat ADCS. Further, to determine the absolute pointing of the camera, LUVCam2 will use optical images of background stars to perform on-orbit plate-solving. 

\section{PAYLOAD OVERVIEW} \label{sec:payload}

The LUVCam2 payload is allocated a maximum weight of 1 kg and equivalent volume of $\sim$1.4U aboard BRNOsat (Figure \ref{fig:tuna_can}). The payload consists of the optomechanical enclosure, sensor and readout electronics, and heater and thermal control system. In the following sections, we describe in detail each of the payload subsystems for LUVCam2.

\subsection{Detector}

\begin{figure}[ht]
    \centering
    \includegraphics[width=0.9\linewidth]{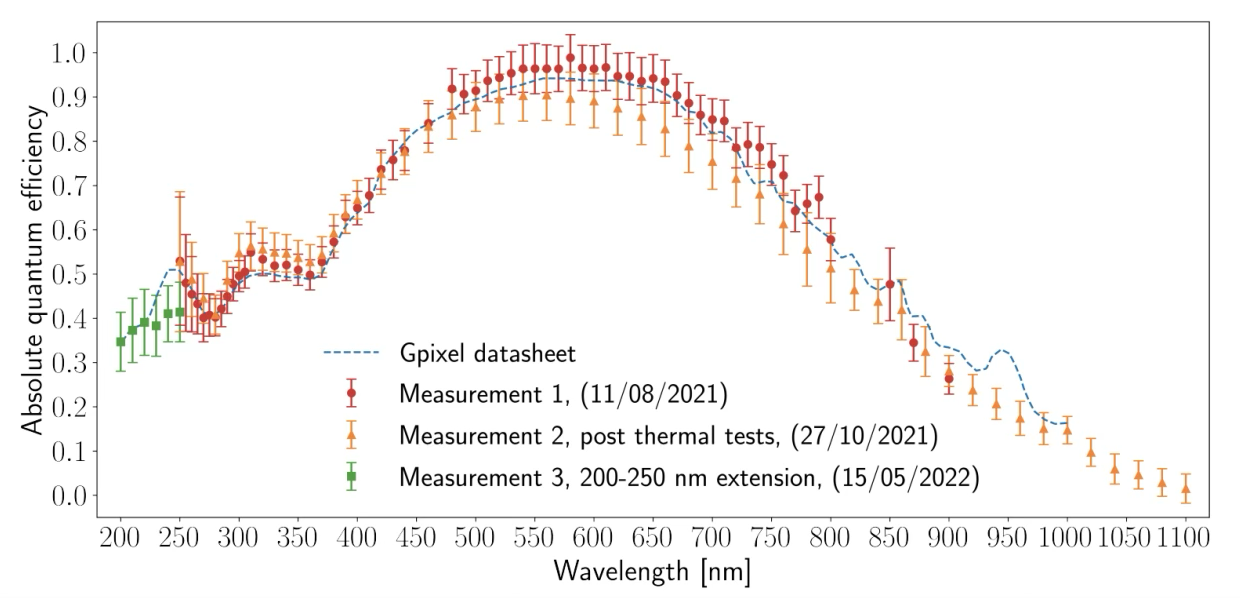}
    \caption{The quantum efficiency measurement of the GSENSE2020BSI sensor using the characterization system at the Dunlap Institute\cite{Gill_2022}. The GSENSE2020BSI is a smaller model of the GSENSE4040BSI, with a sensitivity expected to match that of the GSENSE4040BSI.}
    \label{fig:qe}
\end{figure}

LUVCam2 will integrate the GSENSE4040BSI sensor, a commercial-off-the-shelf scientific CMOS sensor with competitive UV sensitivity. The quantum efficiency of the GSENSE4040BSI has been thoroughly characterized at the Dunlap Institute\cite{Jeram_2024}, and its low read noise and relatively low dark current will enable background-limited observations of scientific targets (Table \ref{tab:gsense}). A similar QE curve for a smaller model of our sensor, the GSENSE2020BSI, has been measured at the Dunlap Institute and is presented in Figure \ref{fig:qe}.

\begin{table}[ht]
    \caption{The technical specification and justifications for the science sensor (GSENSE4040BSI).}
    \vspace{3mm}
    \centering
    \small
    \begin{tabular}{|p{3cm}|p{3cm}|p{9cm}|}
    \hline
    \textbf{Parameter} & \textbf{Specification} & \textbf{Justification} \\
    \hline
    Peak Quantum Efficiency (QE) & 90\% @ 580 nm & The high QE maximizes the sensitivity of the observatory. For UV imaging cases however this puts stricter requirements on the red-leak attenuation. \\
    \hline
    Peak UV Quantum Efficiency (QE) (200--300nm) & 55\% & The relatively high UV QE dramatically expands science opportunities into the UV. \\
    \hline
    Detector Size & $36.8 \times 36.8$ mm & \multirow{3}{9cm}{The large format and pixels of the detector enable achievable optical designs and high-resolution images.} \\
    \cline{1-2}
    Imaging Pixel Array & $4096 \times 4096$ & \\
    \cline{1-2}
    Pixel size & $9 \times 9$ microns & \\
    \hline
    Read Noise & 2.3 $e^{-}$ (High gain) & \multirow{2}{9cm}{The low read noise and relatively low dark current allow for both fast imaging as well as background-limited observations.} \\
    \cline{1-2}
    Dark Current (-15C) & 0.10 $e^{-}$/pix/s & \\
    \hline
    Full Well Capacity & 39,000 $e^{-}$ (Low gain) & \multirow{2}{9cm}{The well capacity and low read noise yields dynamic range sufficient for a variety of science goals.} \\
    \cline{1-2}
    Analog-to-Digital Converter & 12-bit on two (high/low) gain channels & \\
    \hline
    Full Frame Rate & 48 Hz & This extends capabilities to high-speed photometry. Substantially faster readout is possible for smaller Regions-of-Interest. \\
    \hline
    \end{tabular}
    \label{tab:gsense}
\end{table}

\subsection{Optical Design}

\subsubsection{Performance Requirements}

A combination of spacecraft constraints, science mission goals, and operational requirements determines the performance specifications of LUVCam2's optical design:
\begin{itemize}
    \item LUVCam2 will image background stars in the optical field to perform on-orbit plate-solving, since it is denser than the UV background field. This constrains the minimum FOV for LUVCam2's optical field.
    \item A guiding loop will pass centroid positions as feedback to ADCS to maintain pointing. This requires sufficient sensitivity in the optical field for obtaining accurate centroid positions under quick exposures.
    \item In the absence of stable pointing, spacecraft jitter significantly degrades the quality of the PSF. We thus constrain the PSF size to be smaller than the noise induced by spacecraft jitter for science exposures.
    \item A minimum of one additional UV bright star must be detectable in LUVCam2's field of view for each flare observation to perform differential photometry on the science target.
    \item Physical payload requirements limit the maximum total track length to 105\,mm and primary mirror diameter to 80\,mm.
\end{itemize}

To limit read noise and prevent streaking of the source over several pixels, we limit exposures used for on-sensor plate-solving to be 100\,ms. 
Given the significant level of spacecraft jitter that reduces image quality, we bound the image quality of our optics to perform no worse than the associated spacecraft random pointing error of 10$''$ for a 100\,ms exposure\footnote{As measured for the ADCS on GRBBeta, currently in-orbit.}. 
A clear aperture of 80\,mm maximizes the available payload volume on the +Z free panel of BRNOsat. A focal length of 300\,mm (chosen to minimize aberrations while maximizing mirror separation in the allocated payload volume) sets the plate scale of the detector to be 6.19$''$/pix. This corresponds to a physical spot size of 14.54\,$\mu$m ($\sim$1.6 pixel radius) on the detector, a requirement that is imposed across both the UV and optical bands. 
To minimize production costs and complexity of our optical design, we physically divide the LUVCam2 field of view into UV- and optical-sensitive portions. The field of view is partitioned $\sim\frac{2}{3}$ and $\sim\frac{1}{3}$ in area between the UV and optical bands, with the minimum optical field of view driven by the requirement for plate-solving. 
The performance requirements for the LUVCam2 telescope are summarized in Table \ref{tab:requirements}.

\begin{table}[ht]
\caption{Performance requirements for the optical system of LUVCam2.}
\label{tab:requirements}
\vspace{2mm}
\centering
\begin{tabular}{|l|l|}
\hline
\rule[-1ex]{0pt}{3.5ex}  \textbf{Parameter} & \textbf{Specification} \\
\hline
\rule[-1ex]{0pt}{3.5ex}  Aperture & 80\,mm\\
\hline
\rule[-1ex]{0pt}{3.5ex}  Total track length & 105\,mm\\
\hline
\rule[-1ex]{0pt}{3.5ex}  Focal length & 300\,mm\\
\hline
\rule[-1ex]{0pt}{3.5ex}  Plate scale & 6.19''/pix   \\
\hline
\rule[-1ex]{0pt}{3.5ex}  Field of view & UV: 11\,deg$^2$; Optical: 5\,deg$^2$\\
\hline
\rule[-1ex]{0pt}{3.5ex}  Sensitivity (60s exposure) & UV: 18.7 AB mag; Optical: 19.8 AB mag\\
\hline
\rule[-1ex]{0pt}{3.5ex}  Image quality & 1.6 pixel radius at 80\% enclosed energy in optical/UV  \\
\hline
\rule[-1ex]{0pt}{3.5ex}  Passbands & UV: 250--350\,nm; Optical: 410--550\,nm\\
\hline
\rule[-1ex]{0pt}{3.5ex}  Pixel scale \& size & 9\,$\mu$m pixel; $4096\times4096$ pixels\\
\hline
\rule[-1ex]{0pt}{3.5ex}  Active area of sensor used & 21$\times$21\,mm; $2327\times2327$ pixels\\
\hline
\end{tabular}
\end{table}

\subsubsection{Design Specifications}

\begin{figure}[ht]
    \centering
    \includegraphics[width=0.9\linewidth]{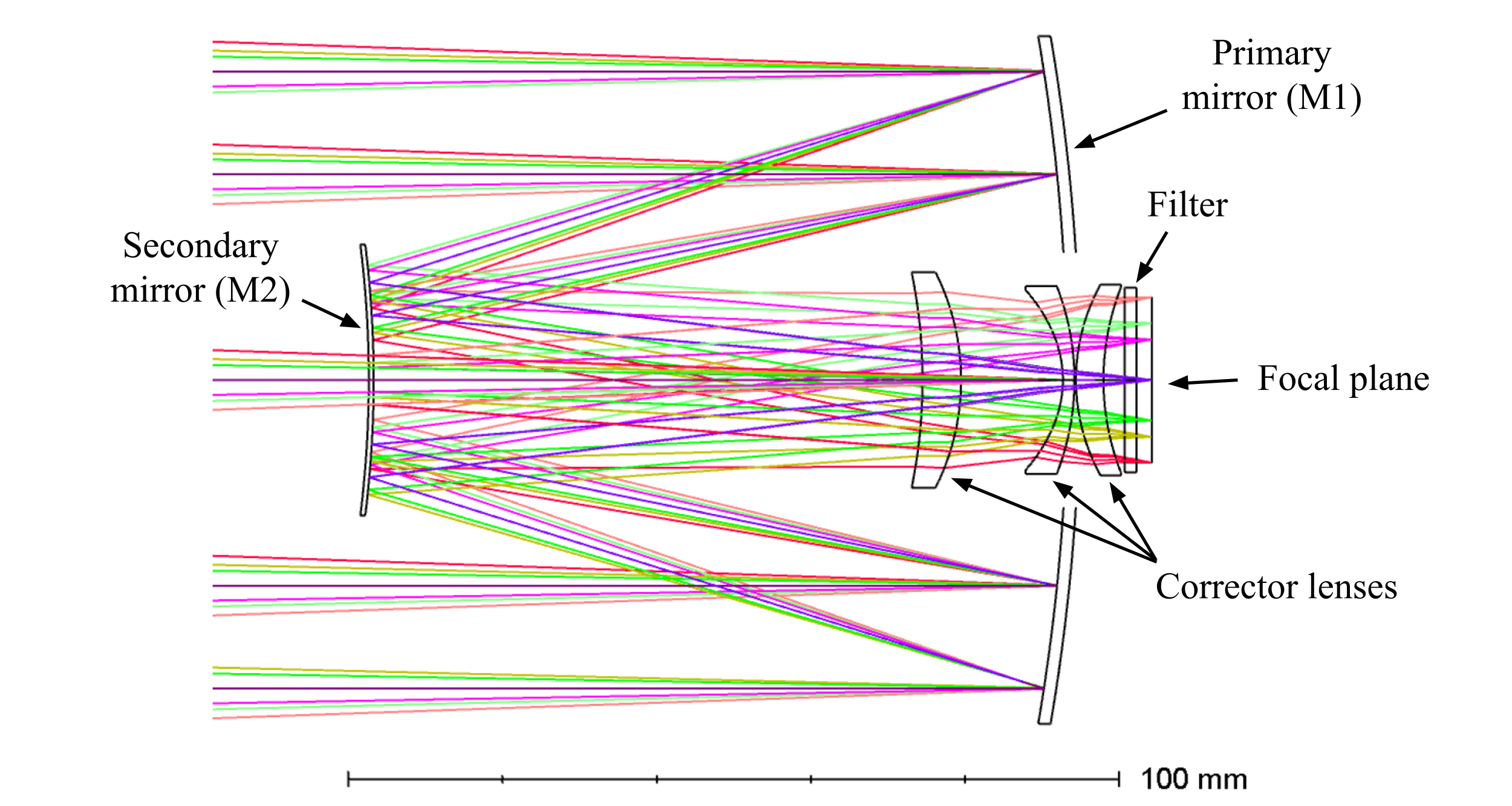}
    \caption{Ray trace of the optical design for LUVCam2, shown in the Y-Z plane.} 
    \label{fig:optics}
\end{figure}

The requirements are met by a Cassegrain design with three corrector lenses (Figure \ref{fig:optics}). The primary mirror and lens surfaces are spherical, and the secondary mirror is hyperbolic. The lens profiles are optimized to minimize the PSF size across the full field.

\begin{figure}[ht]
    \centering
    \includegraphics[width=0.7\linewidth]{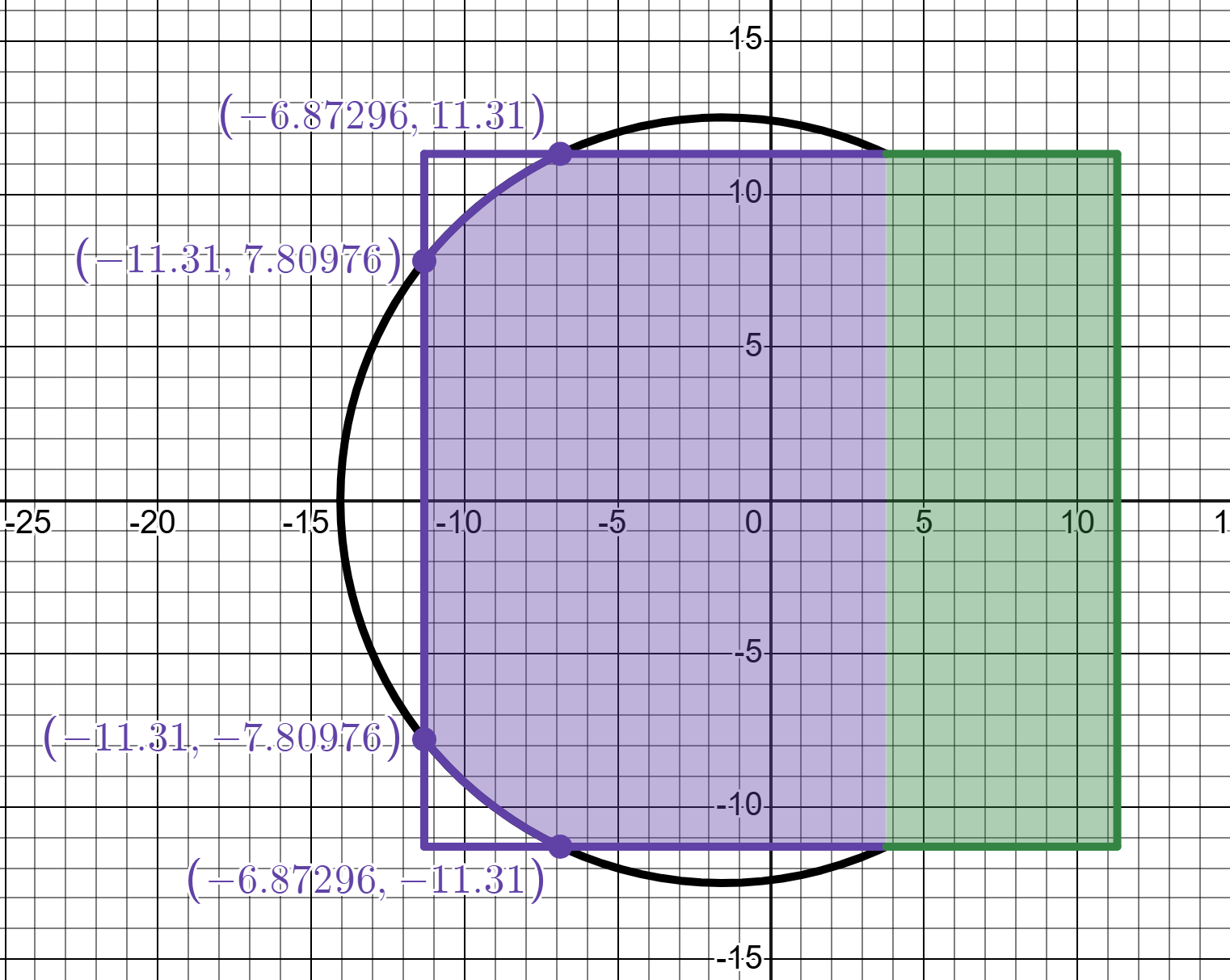}
    \caption{Physical arrangement of the split-field filter for LUVCam2. The green shaded region denotes the optical band (410--550\,nm), and the purple shaded region denotes the UV band (250--350\,nm). The black outline is the physical extent of the UV filter, assuming a circular filter with a 25mm diameter that is cut such that the diced edge corresponds to a 4$^\circ$ FOV in the filter-plane. Axes are in units of millimetres, in the plane of the filter.}
    \label{fig:filter}
\end{figure}

An aperture stop is placed at the primary mirror to ensure that the image is unvignetted across the full 4$^\circ\times4^\circ$ field. The UV and optical filters are held in the same plane above the detector, such that the field of view is physically split into a UV-sensitive and optical-sensitive portion. Figure \ref{fig:filter} shows the split-field filter arrangement between the UV and optical filters. A customized mechanical retainer will mount the UV and optical filters side-by-side, minimizing the detector dead zone directly below the filter boundaries. We will use a combination of UV fused silica and calcium fluoride components to optimize for broadband performance while keeping manufacturing costs low. The expected average optical throughput of the system (excluding filter transmission) is $\sim$75\% across both bands. 

\subsection{Electronics}

\begin{figure}[ht]
    \centering
    \includegraphics[width=0.98\linewidth]{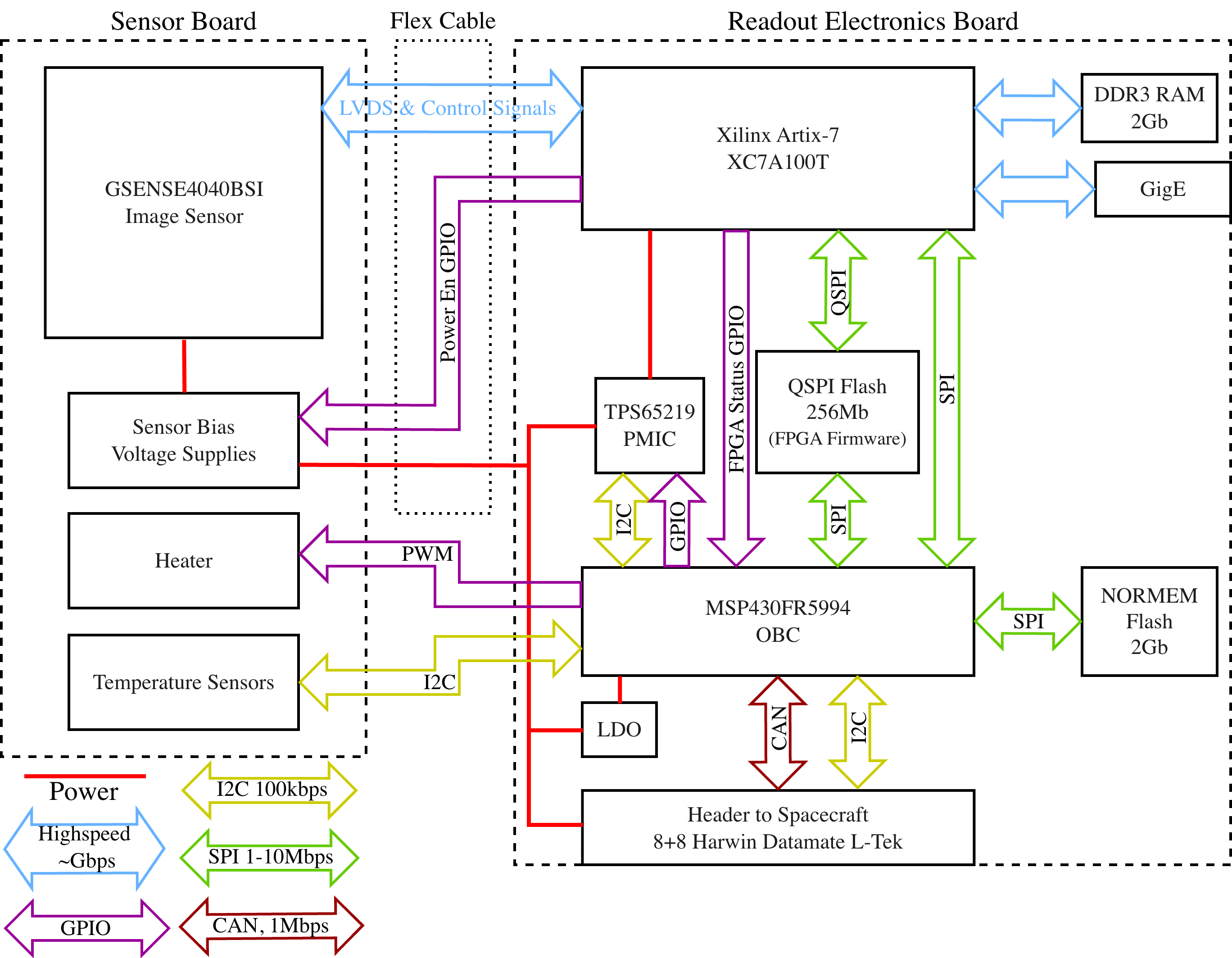}
    \vspace{3mm}
    \caption{LUVCam2 Electrical Diagram. Left side: Sensor Board. Right side: Readout Electronics (RE) Board.}
    \label{fig:electronics}
\end{figure}

The electronics builds on the space-heritage achieved by LUVCam1. The architecture remains largely unchanged. Like its predecessor, LUVCam2 can be separated into two main components: (1) the Readout Electronics board and (2) the Sensor board (Figure \ref{fig:electronics}).

\subsubsection{Readout Electronics}
The Readout Electronics (RE) board features the Xilinx Artix-7 FPGA used on LUVCam1. The FPGA handles power sequencing, sensor training, exposure and readout. It receives commands from the OBC to execute these tasks. The same chip, the MSP430FR5994, was used as GRBBeta's OBC. The RE board essentially incorporates GRBBeta's OBC into LUVCam1's readout electronics to make LUVCam2 an independent payload. LUVCam1 previously relied on GRBBeta's OBC to send the imaging commands and also store the images. 

The CMOS imaging sequence is as follows: (1) The OBC sends the sensor boot command to the FPGA which power sequences the bias voltages. (2) The OBC sends the sensor train command which matches the delays in the LVDS lines between the sensor and the FPGA. (3) The OBC sets the exposure time register on the FPGA, triggers exposure, and reads the data out. 

The RE board interfaces with the Sensor board via two flex cables. One is an 80-pin flex cable that handles the high speed differential lines. To reduce cross-talk, there is a ground line between each pair of lines allocated as a differential pair. The other is a 50-pin cable that handles the low speed digital lines. In both cables, not all pins are utilized. Some lines are left unused on purpose for future compatibility with other sensors. In the same spirit of compatibility, the TPS65219 PMIC output is configurable by setting its registers via I2C. This allows us to change the bank voltages of the FPGA which allows us to be flexible with different communication protocols that a future sensor could have. Different protocols require different voltages.

\subsubsection{Sensor Board}
While the RE board can remain the same between different sensors, the Sensor board must be modified for each sensor-- which was the case going to LUVCam2 from LUVCam1. Fortunately, this board is much simpler compared to the RE board. It consists only of the sensor socket, bias voltage supplies, temperature sensors, and the heater. The sensor socket and bias voltage supplies change per sensor. 

While both missions use the GSENSE4040 detector, LUVCam1 uses the frontside illuminated (FSI) model whereas LUVCam2 uses the backside illuminated (BSI) model. The footprint is the same between both models, but otherwise, changes included: different main clock supply scheme, different voltage bias levels, and some pins having different functionality. These were minor changes but paired also with using a different connector, a new board was necessary. With the new design, we length-matched the LVDS lines to within 1 mm, added mounting holes for the heater, and improve noise filtering on the bias voltage supplies. 

The current Sensor board model used for testing utilizes the same Molex FPC connectors used on the detector board. For flight, this will be removed in favor of a flex-rigid design. The flight Sensor board will be fabricated with the cables already mounted on some of the PCB's layers. This eliminates a point of failure, but also lets us tailor the cable length to be just right. 

\subsection{Mechanical \& Thermal}

\begin{figure}
    \centering
    \includegraphics[width=0.4\linewidth]{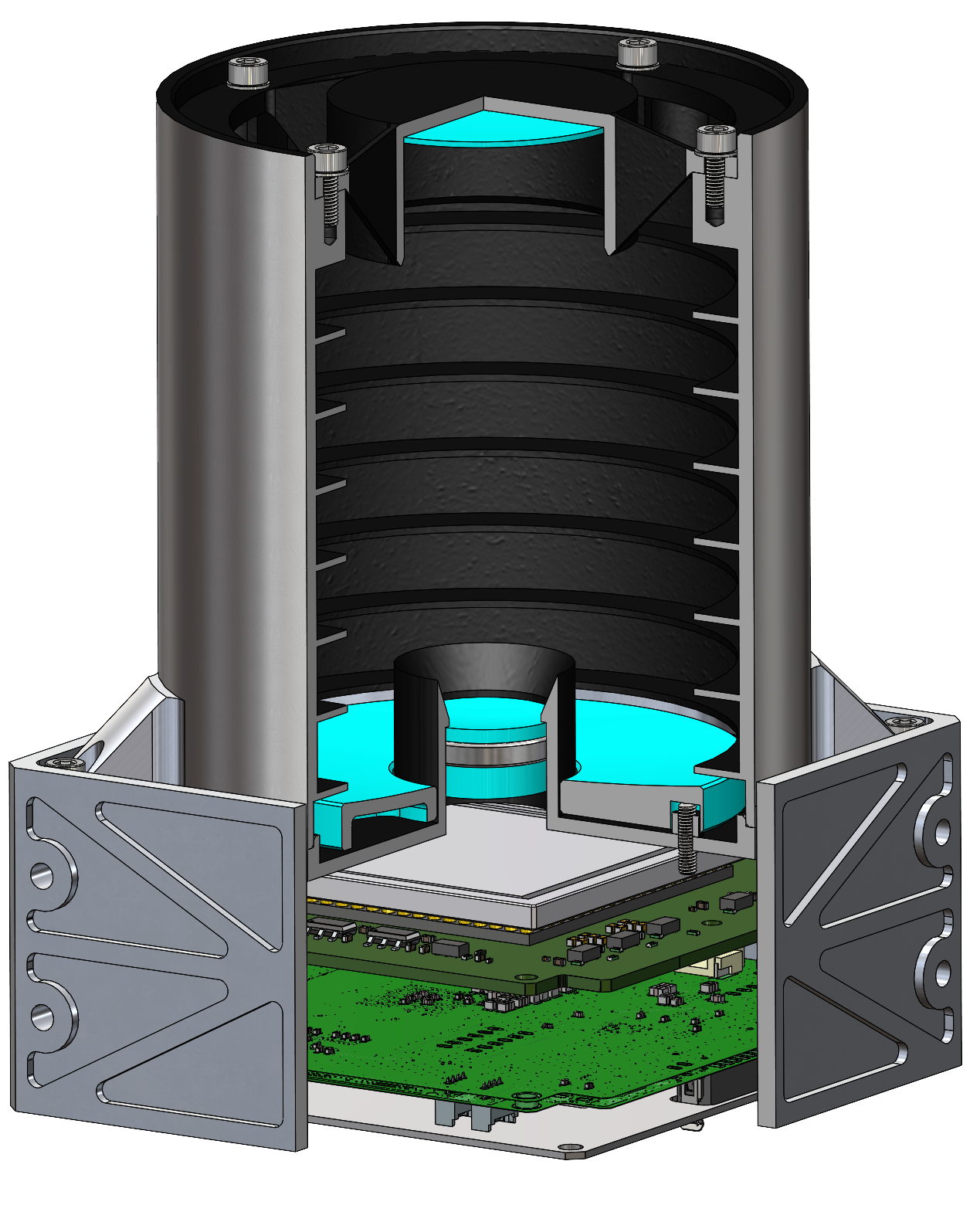}
    \caption{Cross-section view of the mechanical layout for the LUVCam2 payload.}
    \label{fig:mechanical}
\end{figure}

\begin{figure}
\centering
\begin{minipage}{.5\textwidth}
  \centering
  \includegraphics[width=0.8\linewidth]{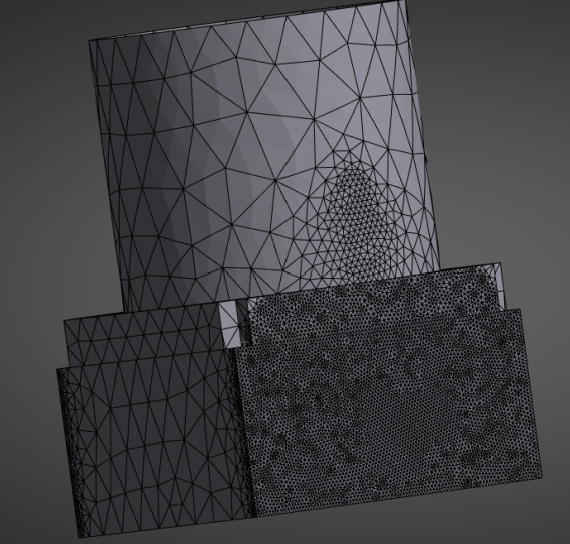}
\end{minipage}
\begin{minipage}{.475\textwidth}
  \centering
  \includegraphics[width=0.8\linewidth]{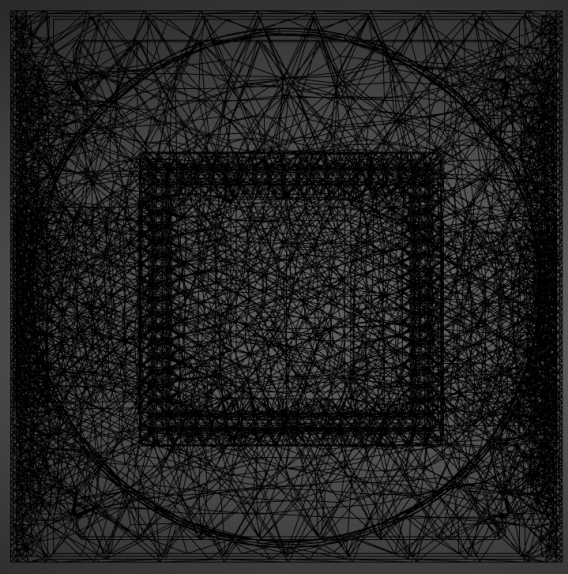}
\end{minipage}
\vspace{1mm}
\caption{The exterior (left) and interior (right) meshed LUVCam2 model used for transient thermal analysis.}
\label{fig:thermal}
\end{figure}

Preliminary mechanical integration for the LUVCam2 payload is shown in Figure \ref{fig:mechanical}. The LUVCam2 thermal system will use a Kapton patch heater for active thermal control of the detector assembly. Aluminized FEP tape will be placed on free faces of the LUVCam2 enclosure to provide passive heat rejection. Preliminary static and transient thermal analysis for modelling the radiative environment of LUVCam2 in low-Earth orbit has been performed (Figure \ref{fig:thermal}). To maintain a nominal system operating temperature of $-20 \pm 5^{\circ}$\,C, we require a 1\,W 5\,cm$\times$5\,cm heater and 90\,cm$^2$ radiator ($\sim$2 faces of the LUVCam2 enclosure).

\section{CURRENT DEVELOPMENT STATUS} \label{sec:current}
The optical design has been finalized, and we are currently evaluating options for manufacturing the components. Based on initial quotes from bidders, the lead time for all components are between 8--10 weeks, with expected delivery by late 2026. Preliminary optical assembly and integration will begin in early 2027. 

The GSENSE4040BSI sensor has been integrated onto the Sensor Board and first light has been achieved in-lab. Gain and read noise tests are to follow. Additionally, a rigid-flex PCB for the Sensor board to incorporate the flex cables onto its PCB is being made. 

The thermal system and mechanical design will be finalized prior to payload integration in early 2027. 

\section{CONCLUSION} \label{sec:conclusion}

This work overviews the science goals, operational capabilities, and system architecture of the second Little UltraViolet Camera (LUVCam2). LUVCam2 is a $\sim$1.4U UV-optical telescope that will observe the UV flares of low-mass stars, providing dual-colour photometry and constraints on UV-optical flare energies of M-dwarfs. The LUVCam2 mission is also a tech demonstration of its novel low-cost, general CMOS image sensor electronics platform. For this iteration of the LUVCam electronics board, the camera integrates the GSENSE4040BSI, a commercial off-the-shelf CMOS sensor by GPixel. LUVCam2 will fly aboard BRNOsat, a Czech 3U CubeSat mission with launch manifested for late 2027. LUVCam2's payload development is on schedule for payload finalization in early 2027.

\acknowledgments
 
This work is enabled by the Dunlap Institute for Astronomy \& Astrophysics. The Dunlap Institute is funded through an endowment established by the David Dunlap family and the University of Toronto. The LUVCam2 project is part of a long-term collaboration between the Dunlap Institute and Masaryk University in Brno, Czech Republic. NW, FM, and JR thank the support by the Czech Science Foundation (GAČR) project No. 24-11487J.

\bibliography{report}

@INPROCEEDINGS{Gill_2022,
       author = {{Gill}, Ajay S. and {Shaaban}, Mohamed M. and {Tohuvavohu}, Aaron and {Sivanandam}, Suresh and {Abraham}, Roberto G. and {Chen}, Seery and {Drout}, Maria R. and {Lokhorst}, Deborah and {Matzner}, Christopher D. and {Mochnacki}, Stefan W. and {Netterfield}, Calvin B.},
        title = "{A low-cost ultraviolet-to-infrared absolute quantum efficiency characterization system of detectors}",
    booktitle = {X-Ray, Optical, and Infrared Detectors for Astronomy X},
         year = 2022,
       editor = {{Holland}, Andrew D. and {Beletic}, James},
       series = {Society of Photo-Optical Instrumentation Engineers (SPIE) Conference Series},
       volume = {12191},
        month = aug,
          eid = {1219114},
        pages = {1219114},
          doi = {10.1117/12.2627564},
archivePrefix = {arXiv},
       eprint = {2207.13052},
 primaryClass = {astro-ph.IM},
       adsurl = {https://ui.adsabs.harvard.edu/abs/2022SPIE12191E..14G}
}

@INPROCEEDINGS{Jeram_2024,
       author = {{Jeram}, Sarik and {Van-Lane}, Phil and {Gail}, Braden and {Taylor}, Jacob and {Shaaban}, Mohamed and {Gill}, Ajay and {Tohuvavohu}, Aaron and {Sivanandam}, Suresh},
        title = "{Detector characterization of the Little Ultraviolet Camera (LUVCamera)}",
    booktitle = {X-Ray, Optical, and Infrared Detectors for Astronomy XI},
         year = 2024,
       editor = {{Holland}, Andrew D. and {Minoglou}, Kyriaki},
       series = {Society of Photo-Optical Instrumentation Engineers (SPIE) Conference Series},
       volume = {13103},
        month = aug,
          eid = {131031Y},
        pages = {131031Y},
          doi = {10.1117/12.3020473},
       adsurl = {https://ui.adsabs.harvard.edu/abs/2024SPIE13103E..1YJ}
}

@article{
Rimmer_2018,
author = {Paul B. Rimmer  and Jianfeng Xu  and Samantha J. Thompson  and Ed Gillen  and John D. Sutherland  and Didier Queloz },
title = {The origin of RNA precursors on exoplanets},
journal = {Science Advances},
volume = {4},
number = {8},
pages = {eaar3302},
year = {2018},
doi = {10.1126/sciadv.aar3302},
URL = {https://www.science.org/doi/abs/10.1126/sciadv.aar3302},
eprint = {https://www.science.org/doi/pdf/10.1126/sciadv.aar3302}}

@article{Davenport_2012,
   title={MULTI-WAVELENGTH CHARACTERIZATION OF STELLAR FLARES ON LOW-MASS STARS USING SDSS AND 2MASS TIME-DOMAIN SURVEYS},
   volume={748},
   ISSN={1538-4357},
   url={http://dx.doi.org/10.1088/0004-637X/748/1/58},
   DOI={10.1088/0004-637x/748/1/58},
   number={1},
   journal={The Astrophysical Journal},
   publisher={American Astronomical Society},
   author={Davenport, James R. A. and Becker, Andrew C. and Kowalski, Adam F. and Hawley, Suzanne L. and Schmidt, Sarah J. and Hilton, Eric J. and Sesar, Branimir and Cutri, Roc},
   year={2012},
   month=mar, pages={58} }

@ARTICLE{Gershberg_1999,
       author = {{Gershberg}, R.~E. and {Katsova}, M.~M. and {Lovkaya}, M.~N. and {Terebizh}, A.~V. and {Shakhovskaya}, N.~I.},
        title = "{Catalogue and bibliography of the UV Cet-type flare stars and related objects in the solar vicinity}",
      journal = {Astronomy and Astrophysics Supplement},
         year = 1999,
        month = nov,
       volume = {139},
        pages = {555-558},
          doi = {10.1051/aas:1999407},
       adsurl = {https://ui.adsabs.harvard.edu/abs/1999A&AS..139..555G}
}

@article{Boksenberg_1973,
	title = {The ultra-violet sky-survey telescope in the {TD}-{IA} satellite.},
	volume = {163},
	issn = {0035-8711},
	url = {https://ui.adsabs.harvard.edu/abs/1973MNRAS.163..291B},
	doi = {10.1093/mnras/163.3.291},
	urldate = {2025-10-24},
	journal = {Monthly Notices of the Royal Astronomical Society},
	author = {Boksenberg, A. and Evans, R. G. and Fowler, R. G. and Gardner, I. S. K. and Houziaux, L. and Humphries, C. M. and Jamar, C. and Macau, D. and Malaise, D. and Monfils, A. and Nandy, K. and Thompson, G. I. and Wilson, R. and Wroe, H.},
	month = jan,
	year = {1973},
	note = {Publisher: OUP
ADS Bibcode: 1973MNRAS.163..291B},
	pages = {291},
}

@article{Ranjan_2017,
doi = {10.3847/1538-4357/aa773e},
url = {https://doi.org/10.3847/1538-4357/aa773e},
year = {2017},
month = {jul},
publisher = {The American Astronomical Society},
volume = {843},
number = {2},
pages = {110},
author = {Ranjan, Sukrit and Wordsworth, Robin and Sasselov, Dimitar D.},
title = {The Surface UV Environment on Planets Orbiting M Dwarfs: Implications for Prebiotic Chemistry and the Need for Experimental Follow-up},
journal = {The Astrophysical Journal}
}

@article{Rimmer_2021,
author = {Paul B. Rimmer and Samantha J. Thompson and Jianfeng Xu and David A. Russell and Nicholas J. Green and Dougal J. Ritson and John D. Sutherland and Didier P. Queloz},
title ={Timescales for Prebiotic Photochemistry Under Realistic Surface Ultraviolet Conditions},

journal = {Astrobiology},
volume = {21},
number = {9},
pages = {1099-1120},
year = {2021},
doi = {10.1089/ast.2020.2335},
    note ={PMID: 34152196},

URL = { 
    
        https://doi.org/10.1089/ast.2020.2335
    
    

},
eprint = { 
    
        https://doi.org/10.1089/ast.2020.2335
    
    

}
}

@article{Mamonova_2026,
	author = {{Mamonova, E.} and {Kowalski, A. F.} and {Herbst, K.} and {Wedemeyer, S.} and {Werner, S. C.}},
	title = {Young M-dwarfs flare activity model: Towards better exoplanetary atmospheric characterisation},
	DOI= "10.1051/0004-6361/202556844",
	url= "https://doi.org/10.1051/0004-6361/202556844",
	journal = {Astronomy \& Astrophysics},
	year = 2026,
	volume = 705,
	pages = "A165",
}

@article{France_2013,
doi = {10.1088/0004-637X/763/2/149},
url = {https://doi.org/10.1088/0004-637X/763/2/149},
year = {2013},
month = {jan},
publisher = {The American Astronomical Society},
volume = {763},
number = {2},
pages = {149},
author = {France, Kevin and Froning, Cynthia S. and Linsky, Jeffrey L. and Roberge, Aki and Stocke, John T. and Tian, Feng and Bushinsky, Rachel and Désert, Jean-Michel and Mauas, Pablo and Vieytes, Mariela and Walkowicz, Lucianne M.},
title = {THE ULTRAVIOLET RADIATION ENVIRONMENT AROUND M DWARF EXOPLANET HOST STARS*},
journal = {The Astrophysical Journal}
}

@ARTICLE{Gunther_2020,
       author = {{G{\"u}nther}, Maximilian N. and {Zhan}, Zhuchang and {Seager}, Sara and {Rimmer}, Paul B. and {Ranjan}, Sukrit and {Stassun}, Keivan G. and {Oelkers}, Ryan J. and {Daylan}, Tansu and {Newton}, Elisabeth and {Kristiansen}, Martti H. and {Olah}, Katalin and {Gillen}, Edward and {Rappaport}, Saul and {Ricker}, George R. and {Vanderspek}, Roland K. and {Latham}, David W. and {Winn}, Joshua N. and {Jenkins}, Jon M. and {Glidden}, Ana and {Fausnaugh}, Michael and {Levine}, Alan M. and {Dittmann}, Jason A. and {Quinn}, Samuel N. and {Krishnamurthy}, Akshata and {Ting}, Eric B.},
        title = "{Stellar Flares from the First TESS Data Release: Exploring a New Sample of M Dwarfs}",
      journal = {The Astronomical Journal},
         year = 2020,
        month = feb,
       volume = {159},
       number = {2},
          eid = {60},
        pages = {60},
          doi = {10.3847/1538-3881/ab5d3a},
archivePrefix = {arXiv},
       eprint = {1901.00443},
 primaryClass = {astro-ph.EP},
       adsurl = {https://ui.adsabs.harvard.edu/abs/2020AJ....159...60G}
}

@article{Jackman_2022,
author = { James A. G. Jackman and Evgenya Shkolnik and Chase Million and Scott Fleming and Tyler Richey-Yowell and Parke Loyd },
title = { Extending Optical Flare Models to the UV: Results from Comparing of TESS and GALEX Flare Observations For M Dwarfs },
eprint = { Arxiv:2210.15688v1 },
year = { 2022 },
url = { https://arxiv.org/abs/2210.15688 },
}

@INPROCEEDINGS{Ishida_1990,
       author = {{Ishida}, K.},
        title = "{Statistics of Flares Observed for Uv-Ceti Type Stars Yz-Canis Ad-Leonis and Ev-Lacertae at the Okayama Observatory}",
    booktitle = {Flare Stars in Star Clusters, Associations and the Solar Vicinity},
         year = 1990,
       editor = {{Mirzoian}, L.~V. and {Pettersen}, B.~R. and {Tsvetkov}, M.~K.},
       series = {IAU Symposium},
       volume = {137},
        month = jan,
        pages = {43},
       adsurl = {https://ui.adsabs.harvard.edu/abs/1990IAUS..137...43I}
}

@ARTICLE{Gershberg_1972,
       author = {{Gershberg}, R.~E.},
        title = "{Some results of the cooperative photometric observations of the UV Cet-type flare stars in the years 1967-71}",
      journal = {Astrophysics and Space Science},
         year = 1972,
        month = nov,
       volume = {19},
       number = {1},
        pages = {75-92},
          doi = {10.1007/BF00643168},
       adsurl = {https://ui.adsabs.harvard.edu/abs/1972Ap&SS..19...75G}
}

@article{Werner_2024,
	title = {Science with a {Small} {Two}-{Band} {UV}-{Photometry} {Mission} {I}: {Mission} {Description} and {Follow}-up {Observations} of {Stellar} {Transients}},
	volume = {220},
	issn = {0038-6308},
	shorttitle = {Science with a {Small} {Two}-{Band} {UV}-{Photometry} {Mission} {I}},
	url = {https://www.scopus.com/pages/publications/85184246559},
	doi = {10.1007/s11214-024-01048-3},
	number = {11},
	urldate = {2025-10-27},
	journal = {Space Science Reviews},
	author = {Werner, N. and Řípa, J. and Thöne, C. and Münz, F. and Kurfürst, P. and Jelínek, M. and Hroch, F. and Benáček, J. and Topinka, M. and Lukes-Gerakopoulos, G. and Zajaček, M. and Labaj, M. and Prišegen, M. and Krtička, J. and Merc, J. and Pál, A. and Pejcha, O. and Dániel, V. and Jon, J. and Šošovička, R. and Gromeš, J. and Václavík, J. and Steiger, L. and Segiňák, J. and Behar, E. and Tarem, S. and Salh, J. and Reich, O. and Ben-Ami, S. and Barschke, M. F. and Berge, D. and Tohuvavohu, A. and Sivanandam, S. and Bulla, M. and Popov, S. and Chang, Hsiang Kuang},
	month = feb,
	year = {2024},
}

@inproceedings{Daniel_2024,
	title = {{QUVIK} ({Quick} {Ultra}-{VIolet} {Kilonova} surveyor) spacecraft and payload system design overview},
	volume = {13093},
	url = {https://www.spiedigitallibrary.org/conference-proceedings-of-spie/13093/1309306/QUVIK-Quick-Ultra-VIolet-Kilonova-surveyor-spacecraft-and-payload-system/10.1117/12.3021362.full},
	doi = {10.1117/12.3021362},
	urldate = {2025-10-27},
	booktitle = {Space {Telescopes} and {Instrumentation} 2024: {Ultraviolet} to {Gamma} {Ray}},
	publisher = {SPIE},
	author = {Dániel, V. and Werner, N. and Václavík, J. and Jon, J. and Svoboda, P. and Rak, F. and Hriadel, D. and Gromeš, J. and Šošovička, R. and Munz, F. and Jelínek, M. and Steiger, L. and Segiňák, J. and Řípa, J.},
	month = aug,
	year = {2024},
	pages = {33--39},
}

@INPROCEEDINGS{Werner_2022,
       author = {{Werner}, N. and {{\v{R}}{\'\i}pa}, J. and {M{\"u}nz}, F. and {Hroch}, F. and {Jel{\'\i}nek}, M. and {Krti{\v{c}}ka}, J. and {Zaja{\v{c}}ek}, M. and {Topinka}, M. and {D{\'a}niel}, V. and {Grome{\v{s}}}, J. and {V{\'a}clav{\'\i}k}, J. and {Steiger}, L. and {L{\'e}dl}, V. and {Seginak}, J. and {Ben{\'a}{\v{c}}ek}, J. and {Budaj}, J. and {Faltov{\'a}}, N. and {G{\'a}lis}, R. and {Jadlovsk{\'y}}, D. and {Jan{\'\i}k}, J. and {Kajan}, M. and {Karas}, V. and {Kor{\v{c}}{\'a}kov{\'a}}, D. and {Kosiba}, M. and {Krti{\v{c}}kov{\'a}}, I. and {Kub{\'a}t}, J. and {Kub{\'a}tov{\'a}}, B. and {Kurf{\"u}rst}, P. and {Labaj}, M. and {Mikul{\'a}{\v{s}}ek}, Z. and {P{\'a}l}, A. and {Paunzen}, E. and {Piecka}, M. and {Pri{\v{s}}egen}, M. and {Ramezani}, T. and {Skarka}, M. and {Sz{\'a}sz}, G. and {Th{\"o}ne}, C. and {Zejda}, M.},
        title = "{Quick Ultra-VIolet Kilonova surveyor (QUVIK)}",
    booktitle = {Space Telescopes and Instrumentation 2022: Ultraviolet to Gamma Ray},
         year = 2022,
       editor = {{den Herder}, Jan-Willem A. and {Nikzad}, Shouleh and {Nakazawa}, Kazuhiro},
       series = {Society of Photo-Optical Instrumentation Engineers (SPIE) Conference Series},
       volume = {12181},
        month = aug,
          eid = {121810B},
        pages = {121810B},
          doi = {10.1117/12.2629531},
archivePrefix = {arXiv},
       eprint = {2207.05485},
 primaryClass = {astro-ph.IM},
       adsurl = {https://ui.adsabs.harvard.edu/abs/2022SPIE12181E..0BW}
}

@manual{Payne_2026,
  author       = {Payne, A. V. and others},
  title        = {Cosmic Origins Spectrograph Instrument Handbook},
  edition      = {Version 18.0},
  year         = {2026},
  organization = {Space Telescope Science Institute},
  address      = {Baltimore, MD},
}

@ARTICLE{Reid_1997,
       author = {{Reid}, I. Neill and {Gizis}, John E.},
        title = "{Low-Mass Binaries and the Stellar Luminosity Function}",
      journal = {The Astronomical Journal},
         year = 1997,
        month = jun,
       volume = {113},
        pages = {2246},
          doi = {10.1086/118436},
       adsurl = {https://ui.adsabs.harvard.edu/abs/1997AJ....113.2246R}
}

@ARTICLE{Salpeter_1955,
       author = {{Salpeter}, Edwin E.},
        title = "{The Luminosity Function and Stellar Evolution.}",
      journal = {The Astrophysical Journal},
         year = 1955,
        month = jan,
       volume = {121},
        pages = {161},
          doi = {10.1086/145971},
       adsurl = {https://ui.adsabs.harvard.edu/abs/1955ApJ...121..161S}
}

@ARTICLE{Howard_2012,
       author = {{Howard}, Andrew W. and {Marcy}, Geoffrey W. and {Bryson}, Stephen T. and {Jenkins}, Jon M. and {Rowe}, Jason F. and {Batalha}, Natalie M. and {Borucki}, William J. and {Koch}, David G. and {Dunham}, Edward W. and {Gautier}, III, Thomas N. and {Van Cleve}, Jeffrey and {Cochran}, William D. and {Latham}, David W. and {Lissauer}, Jack J. and {Torres}, Guillermo and {Brown}, Timothy M. and {Gilliland}, Ronald L. and {Buchhave}, Lars A. and {Caldwell}, Douglas A. and {Christensen-Dalsgaard}, J{\o}rgen and {Ciardi}, David and {Fressin}, Francois and {Haas}, Michael R. and {Howell}, Steve B. and {Kjeldsen}, Hans and {Seager}, Sara and {Rogers}, Leslie and {Sasselov}, Dimitar D. and {Steffen}, Jason H. and {Basri}, Gibor S. and {Charbonneau}, David and {Christiansen}, Jessie and {Clarke}, Bruce and {Dupree}, Andrea and {Fabrycky}, Daniel C. and {Fischer}, Debra A. and {Ford}, Eric B. and {Fortney}, Jonathan J. and {Tarter}, Jill and {Girouard}, Forrest R. and {Holman}, Matthew J. and {Johnson}, John Asher and {Klaus}, Todd C. and {Machalek}, Pavel and {Moorhead}, Althea V. and {Morehead}, Robert C. and {Ragozzine}, Darin and {Tenenbaum}, Peter and {Twicken}, Joseph D. and {Quinn}, Samuel N. and {Isaacson}, Howard and {Shporer}, Avi and {Lucas}, Philip W. and {Walkowicz}, Lucianne M. and {Welsh}, William F. and {Boss}, Alan and {Devore}, Edna and {Gould}, Alan and {Smith}, Jeffrey C. and {Morris}, Robert L. and {Prsa}, Andrej and {Morton}, Timothy D. and {Still}, Martin and {Thompson}, Susan E. and {Mullally}, Fergal and {Endl}, Michael and {MacQueen}, Phillip J.},
        title = "{Planet Occurrence within 0.25 AU of Solar-type Stars from Kepler}",
      journal = {The Astrophysical Journal Supplement},
         year = 2012,
        month = aug,
       volume = {201},
       number = {2},
          eid = {15},
        pages = {15},
          doi = {10.1088/0067-0049/201/2/15},
archivePrefix = {arXiv},
       eprint = {1103.2541},
 primaryClass = {astro-ph.EP},
       adsurl = {https://ui.adsabs.harvard.edu/abs/2012ApJS..201...15H}
}

@ARTICLE{Mulders_2015,
       author = {{Mulders}, Gijs D. and {Pascucci}, Ilaria and {Apai}, D{\'a}niel},
        title = "{An Increase in the Mass of Planetary Systems around Lower-mass Stars}",
      journal = {The Astrophysical Journal},
         year = 2015,
        month = dec,
       volume = {814},
       number = {2},
          eid = {130},
        pages = {130},
          doi = {10.1088/0004-637X/814/2/130},
archivePrefix = {arXiv},
       eprint = {1510.02481},
 primaryClass = {astro-ph.EP},
       adsurl = {https://ui.adsabs.harvard.edu/abs/2015ApJ...814..130M}
}
\bibliographystyle{spiebib}

\end{document}